\documentclass[11pt, nofootinbib, notitlepage, showpacs, floatfix, tightenlines]{revtex4-1} %longbibliography
\usepackage{amssymb, amsmath, amsfonts, mathrsfs}
\usepackage{footnote}
\usepackage{graphicx}
\usepackage{natbib}
\usepackage{picinpar}
\usepackage{float}

\usepackage{subcaption}

\def\thesection{\arabic{section}}

\begin{document}
\title{Natural inflation with natural number of $e$-foldings}

\author{Matthew Civiletti}\email{matthew.civiletti@qc.cuny.edu}
\author{Brandon Delacruz}
%\email{brandon.delacruz81@qmail.cuny.edu}
\affiliation{Department of Physics, 
Queens College at the City University of New York,
65-30 Kissena Blvd., Queens NY 11367-1597}

\begin{abstract}
We examine natural inflation without the use of the standard slow-roll approximation by considering the number of physical e-folds $\ln (a_eH_e/aH)$. We show that $\tilde{H} = aH \propto \cos(A \phi/2)^{2/A^2} \sin(A \phi/2)$ produces a natural inflationary scenario. This model may be solved exactly, showing that the slow-roll approximation overestimates the tensor-to-scalar ratio by about $13-19 \%$ for $n_s \approx 0.96$ and $50-60$ $e$-folds \footnote{This article is published in \textit{Physical Review D}: 
https://journals.aps.org/prd/abstract/10.1103/PhysRevD.101.043534}.
\end{abstract}

\maketitle

\section{Introduction}
Since the cosmological inflation hypothesis was \cite{guth1981inflationary} proposed, we have witnessed the emergence of precision cosmology. Groundbreaking experiments such as WMAP \cite{bennett2013nine} and Planck \cite{ade2016planck} have given us an unprecedentedly accurate picture of the early universe. This has reduced the number of inflationary models consistent with experimental data, although we are still left with a multitude of potentially valid models. A number of future experiments designed to measure gravitational waves will test large-field models. One such experiment that is scheduled to be launched in the mid-2020s, LiteBIRD \cite{hazumi2019litebird}, should be able to reduce the upper bound on $r$ to $0.002$ at $95\%$ C.L., should no gravitational waves be found. The ground-based experiment QUBIC \cite{tartari2016qubic}, on the other hand, is expected to probe $r$ down to at least $0.05$ at $95\%$ C.L. In this paper, we argue that the slow-roll approximation may be insufficiently accurate in light of future experiments. To that end, we solve a natural inflation scenario exactly, without the use of the standard (potential) slow-roll parameters, and show that this yields errors of the order of $10\%$. We accomplish this by reparametrizing the model so that the number of $e$-folds is defined in a more natural way--i.e., as $\ln a_eH_e/aH$; we then solve for $aH$ for the natural inflationary potential via the Hamilton-Jacobi method. \\
%\subsubsection{Review of the Natural Inflation Scenario}\label{sec_BG}
\indent Let us briefly review the natural inflationary scenario. Natural inflation refers to inflation driven by a potential which is invariant under a shift symmetry $\phi \rightarrow \phi + constant$ \cite{freese1990natural}. In the original manifestation, the potential is 
\begin{equation} 
	V = \Lambda^4 (1 - \cos(\phi/f)),
\label{Wstandard}
\end{equation}
and $\phi$ is the QCD axion \cite{peccei1977cp} field. The status of natural inflation was recently reviewed in \cite{freese2015natural}, within the context of the Planck 2015 \cite{ade2016planck} and BICEP2 results, showing that $r \approx 0.05$ can be achieved at $n_s \approx 0.96$ and $60$ $e$-folds. This requires $\Lambda \sim 10^{16}$ $GeV$ and $f \gtrsim m_{pl}$. This work has been extended to hybrid natural inflation (\cite{ross2016hybrid},\cite{ross2010hybrid},\cite{ross2010hybrid2}) and the so-called extended natural inflation \cite{german2017general}.\\
\indent In this paper we examine a natural inflationary scenario of the form $V = \left( 3 + A^2/2 \right)B \left( 1 - \cos(A \phi) \right) - BA^2$, which is described exactly by $\tilde{H} = aH \propto \cos(A \phi/2)^{2/A^2} \sin(A \phi/2)$, where $\phi$ is the inflaton and $A = 1/f$. Note that $\Lambda = \left[ B \left( 3 + A^2/2 \right)  \right]^{1/4}$. This model has the benefit of being exactly solvable without the use of slow-roll. In addition to providing more accurate results in a natural inflationary scenario, we use this model as an example of how to reparametrize an inflationary model from $V$ to $\tilde{H}$. Further, this type of model is related to the so-called constant-roll inflation, in which $\ddot{\phi}/ H\dot{\phi}$ is constant. This condition may lead to natural inflation with a negative cosmological constant, which is discussed in \cite{motohashi2015inflation}. The inflationary effects of a negative cosmological constant were discussed in \cite{mithani2013inflation}, where it is shown that it generally leads to instabilities. In our model, however, the negative cosmological constant ($-BA^2$) is two orders of magnitude smaller than the value of the potential at Hubble crossing for the entire relevant parameter space. Hence, our model should produce nearly exact solutions to the standard natural inflation model $V = \Lambda^4 (1 - \cos(\phi/f))$.  \\
\indent In this paper, we first review the slow-roll approximation. Then in Section \ref{Natural} we introduce the reparametrization from $V$ to $\tilde{H}$, and explain its relation to the more natural definition of the number of $e$-folds, $\ln a_eH_e/aH$. Finally, we discuss our analytical and numerical results in Section \ref{sec_Results}.

\section{The Slow-Roll Paradigm}
The typical method by which inflationary models are solved is by means of the slow-roll approximation. We first define the Hubble Slow-Roll Parameters (HSRPs), which are defined in general as \cite{liddle1994formalizing}
\begin{equation} \label{HSRdef}
^n \beta_H = \bigg[ \prod_{i=1}^{n} \left( -\frac{d \ln H^{(i)}}{d \ln a} \right) \bigg]^{\frac{1}{n}} = 2 \left( \frac{{H_{,\phi}}^{n-1} H^{\left( n+1\right)}}{H^n}\right)^\frac{1}{n},
\end{equation}
where $H^{\left( n \right)} \equiv \frac{d^{n} H}{d \phi^{n}}$ and $H_{,\phi} \equiv \frac{d H}{d \phi}$, and we separately define $^0\beta_H \equiv \epsilon = 2 \frac{H_{, \phi}^2}{H^2}$. We use $\frac{m_{pl}^2}{8 \pi} = 1$ throughout. The first three terms of this hierarchy are
%\equiv H_{, \phi...\phi} = H_{, \phi,n+1} \equiv \frac{d^{n+1} H}{d \phi^{n+1}}$ such that $n+1$ is the number of $\phi$s in the subscript $\phi...\phi$; and, we separately define $^0\beta_H \equiv \epsilon = 2 \frac{H_{, \phi}^2}{H^2}$. We use $\frac{m_{pl}^2}{8 \pi} = 1$ throughout. The first three terms of this hierarchy are
\begin{equation*}
^0\beta_H \equiv \epsilon = 2 \left( \frac{H_{, \phi}}{H} \right)^2; \hspace{0.3 cm}
^1\beta_H \equiv \eta = 2 \frac{H^{\left( 2 \right)}}{H}; \hspace{0.3 cm}
^2\beta_H \equiv \xi = 2 \left( \frac{H_{, \phi} H^{\left( 3 \right)}}{H^2} \right)^\frac{1}{2}.
%^3\beta_H \equiv \sigma = 2 \left( \frac{H_{, \phi}^2 H_{, \phi, 4}}{H^3} \right)^\frac{1}{3}. 
\end{equation*}
These may be written in terms of another set of parameters which are wholly functions of the potential; we call these the Potential Slow-Roll Parameters (PSRPs), the hierarchy of which is given by
\begin{equation} \label{PSRdef}
^n \beta_V = \frac{d \ln V}{d \phi} \bigg[ \prod_{i=1}^{n} \left( \frac{d \ln V^{(i)}}{d \phi} \right) \bigg]^{\frac{1}{n}} = \left( \frac{{V_{,\phi}}^{n-1} V^{\left( n+1\right)}}{V^n}\right)^\frac{1}{n}.
\end{equation}
Likewise, the first three terms of this hierarchy are
\begin{equation*}
^0\beta_V \equiv \epsilon_v = \frac{1}{2} \left( \frac{V_{, \phi}}{V} \right)^2; \hspace{0.3 cm} 
^1\beta_V \equiv \eta_v = \frac{V^{\left( 2 \right)}}{V}; \hspace{0.3 cm} 
^2\beta_V \equiv \xi_v =  \left( \frac{V_{, \phi} V^{\left( 3 \right)}}{V^2} \right)^\frac{1}{2}.
%^3\beta_V \equiv \sigma_v = \left( \frac{V_{, \phi}^2 V_{, \phi, 4}}{V^3} \right)^\frac{1}{3}.
\end{equation*}

%The lowest order slow-roll parameters are related via \cite{liddle1994formalizing}
%\begin{equation} \label{lowestSRrelation}
%\epsilon_v = \epsilon \left( \frac{3 - \eta}{3 - \epsilon} \right)^2; \hspace{0.3 cm}
%\eta_v = \frac{d \eta}{d \phi} \frac{\sqrt{2 \epsilon}}{3 - \epsilon} + \frac{3 - \eta}{3 - \epsilon} \left( \eta + \epsilon \right). \\
%\end{equation}

From \cite{liddle1994formalizing, stewart1993more}, we employ the next-to-leading order terms for the scalar spectral index and tensor-to-scalar ratio. These are
\begin{equation}\label{HTildeCosmoParameters}
\begin{aligned}
r &= 16 \epsilon \left[ 1 + 2C (\epsilon - \eta) \right], \\      
n_s &= 1 - 4 \epsilon + 2 \eta - \left( 5 - 3C \right) \epsilon^2 \\
    & - \frac{1}{4} \left( 3 - 5C \right) \left( 2 \eta - 4 \epsilon \right) \epsilon  + \frac{1}{2} \left( 3 - C \right) \xi^2, 
\end{aligned}
\end{equation}
where $C = 4\left( \ln 2 + \gamma \right) -5$ and $\gamma$ is Euler's constant. 
%From Equation \ref{HTildeCosmoParameters}, we may derive the first order relations for the first order slow-roll parameters $\epsilon \approx \epsilon_v$ and $\eta \approx \eta_v - \epsilon_v$. 

One may expand the HSRPs in terms of the PSRPs, as discussed in \cite{liddle1994formalizing}; substituting these into Equation \ref{HTildeCosmoParameters}, we obtain results for $n_s$ and $r$ to next to leading order. These are

\begin{equation}\label{SRCosmoParameters}
\begin{aligned}
r_{SR} &= 16 \epsilon_v \left[ 1 + 2 \left( 2 \epsilon_v - \eta_v \right)\left( C - 1/3 \right) \right], \\      
n_{s{SR}} &= 1 - 6 \epsilon_v + 2 \eta_v + \left( \frac{44}{3} - 6C \right) \epsilon_v^2 \\
        & + \frac{2}{3} \eta_v^2 + \epsilon_v \eta_v \left( 4C - 14 \right) + \xi_v^2 \left( \frac{13}{6} - \frac{C}{2}\right). \\
\end{aligned}
\end{equation}

Further, we may compute the curvature perturbations via \cite{lidsey1997reconstructing} to obtain, to first order for simplicity,
\begin{equation} \label{CurvPert}
P_R^{1/2} = \frac{H}{2^{3/2} \pi \sqrt{\epsilon}}.
\end{equation}

%\begin{equation} \label{CurvPert}
%P_R^{1/2} = \frac{H}{2^{3/2} \pi \sqrt{\epsilon}} \left( 1 - \frac{\epsilon}{2} \left( C - 1 \right) + \frac{\eta}{4} \left( C - 3 \right) \right).
%\end{equation}

We may approximate this in slow-roll as
\begin{equation}\label{SRCurvPert}
P_{R_{SR}}^{1/2} = \frac{V^{3/2}}{\sqrt{12} \pi V_{,\phi}}.
\end{equation}

%The Lyth bound \cite{lyth1997would} is 
%\begin{equation} \label{CurvPert}
%V \approx 4.5r \times 10^{-8}
%\end{equation}

%; these are 
%\begin{equation}\label{HTildeCosmoParameters}
%\begin{aligned}
%	r &= \frac{25}{2} \epsilon \left(1 - 2c (\eta - \epsilon) \right) = \frac{25}{2} \epsilon_v \left(1 + 2 \left( c - \frac{1}{3} \right)  \left(2 \epsilon_v - \eta_v) \right) \right), \\
%	n_s &= 1 - 4 \epsilon + 2\eta - 2(1 + c) \epsilon^2 
%- \frac{1}{2}(3 - 5c)\eta \epsilon 
%+ \frac{1}{2}(3 - c)\xi^2 \\
%&=  1 - 6 \epsilon_v + 2\eta_v +\frac{1}{3} (44 - 18c) \epsilon_v^2 
%+ (4c - 14) \eta_v \epsilon_v + \frac{2}{3} \eta_v^2 
%+ \frac{1}{6}(13 - 3c)\xi_v^2, \\
%	\end{aligned}
%\end{equation}
%where $c = 4\left( \ln 2 + \gamma \right)$ and $\gamma$ is Euler's constant.

%%%%%		\alpha*S^4 Term

\section{Natural $e$-folds}\label{Natural}
In the standard slow-roll scenario, one computes the amount of inflation, parametrized via the number of $e$-folds, via the approximation
\begin{eqnarray*}
	N = \int_a^{a_e} \frac{d a}{a} = \int_{\phi_e}^{\phi} \frac{d \phi}{\sqrt{2 \epsilon}} \approx \int_{\phi_e}^{\phi} \frac{d \phi}{\sqrt{2 \epsilon_v}}.
\end{eqnarray*}
The horizon problem, however, can be solved if the comoving Hubble radius $\left( aH \right)^{-1}$ decreases during inflation by a factor of $\frac{\left(a_{0}H_{0}\right)^{-1}}{\left(a_{e}H_{e}\right)^{-1}} = \frac{\left(a_{0}H_{0}\right)^{-1}}{\left(a_{eq}H_{eq}\right)^{-1}} \frac{\left(a_{eq} H_{eq}\right)^{-1}}{\left(a_{e}H_{e}\right)^{-1}} = \sqrt{\frac{a_0}{a_{eq}}} \frac{a_{eq}}{a_{e}} \sim 10^{26}$, where the subscript ``eq" refers to matter-radiation equality, ``e" refers to the end of inflation, and ``0" refers to today. We therefore require $\ln a_eH_e/aH \approx \ln 10^{26} \approx 60$ $e$-folds. If we define $\tilde{N} \equiv \ln a_eH_e/aH$, we may write this as \footnote{In the last step, we use $\dot{\phi} = -2 H_{, \phi}$.}
\begin{equation}\label{nTILDE}
	\tilde{N} = \int_a^{a_e} \frac{da}{a} + \int_H^{H_e} \frac{dH}{H} = \int_\phi^{\phi_e} \left( \frac{a_{, \phi}}{a} + \frac{H_{, \phi}}{H} \right)  d\phi = \int_\phi^{\phi_e} \left( \frac{-1}{\sqrt{2 \epsilon}} + \sqrt{\frac{\epsilon}{2}} \right)  d\phi,
\end{equation}
which we refer to as the number of physical $e$-folds. $\tilde{N}$ is a simpler and more natural way to define the $e$-folds produced by inflation, as discussed more extensively in \cite{liddle1994formalizing}. Recent research has expanded upon this work. In \cite{chongchitnan2016inflation}, \cite{chongchitnan2017inflationary} and {\cite{chongchitnan2017reheating}, it is shown that one can solve inflationary models without the use of slow-roll by computing inflationary observables from an explicit expression for $\tilde{H} \equiv aH$. This allows one to compute the number of $e$-folds without integration, the relationship between $N$ and the physical $e$-folds $\tilde{N}$ being 
\begin{equation}\label{efolds}
	\tilde{N} = N + \ln H_e/H,
\end{equation}
where $\ln H_e/H < 0$. We may then write $\tilde{N}$ in terms of $\tilde{H}$ as
\begin{equation}\label{nTILDEHTilde}
	\tilde{N} =   \int_\phi^{\phi_e}  \frac{\tilde{H}_{, \phi}}{\tilde{H}} d\phi,
\end{equation} 
which we may then compare to Equation \ref{nTILDE} to determine that 
\begin{equation}\label{e1NTilde}
	 \frac{\tilde{H}_{, \phi}}{\tilde{H}} =  \frac{-1}{\sqrt{2 \epsilon}} + \sqrt{\frac{\epsilon}{2}}.
\end{equation}
It is useful to assign this quantity and higher order ones a label. Thus we introduce 
\begin{equation}\label{E1E2def}
	e_1 \equiv  \frac{\tilde{H}_{, \phi}}{\tilde{H}}; \hspace{0.3 cm}
	e_2 \equiv \frac{\tilde{H}^{\left( 2 \right)}}{\tilde{H}}; \hspace{0.3 cm}
	e_3 \equiv \frac{\tilde{H}^{\left( 3 \right)}}{\tilde{H}},
\end{equation}
which, since inflation is defined as $\frac{d}{dt} \left(1/aH \right) <0$, allows one to also define inflation as \footnote{Via Equation \ref{e1NTilde}, this is equivalent to $\epsilon < 1$.} $\frac{d}{dt} \left(aH \right)>0 \Rightarrow e_1 < 0$, assuming that $\dot{\phi}<0$. We can solve Equation \ref{e1NTilde} for $\epsilon$, and subsequently we can also solve for $\eta$ to find that 
%and, using the relationship $\frac{a_{, \phi \phi}}{a} = \frac{1- \epsilon + \eta}{2 \epsilon}$, we can also solve for $\eta$; we find that 
\begin{equation}\label{EpsilonEta}
    \begin{aligned}
	\epsilon &=  e_1^2 + e_1 \sqrt{e_1^2 +2 } + 1, \\
	\eta &= \frac{\epsilon \left(2 e_2 + 3 \right) -1}{\epsilon + 1}.
	\end{aligned}
\end{equation}
In order to connect $\tilde{H}$ to the potential, we use the Hamilton-Jacobi equation
\begin{equation}\label{HJ}
	V =  3H^2 - 2 H_{, \phi}^2,
\end{equation}
which, if solvable for $H$, may be used in conjunction with Equation \ref{e1NTilde} to solve for $e_1$. We obtain the rather uninviting differential equation\footnote{This equation is clearly not valid for $H_{,\phi} = 0$; in this case, $H = constant$ and hence $\tilde{N} = N$ and $\tilde{H}$ is exponential.} 
\begin{equation}\label{HTILDEfromH}
	 \tilde{H} \propto exp \left( \int e_1 d \phi \right) \propto exp \left( \int \frac{-H^2/2 + H_{, \phi}^2}{HH_{, \phi}} d \phi \right). 
\end{equation}
Although analytical solutions are impractical or impossible to obtain for general potentials, we can solve Equations \ref{HJ} and \ref{HTILDEfromH} in the natural inflationary scenario with a negative cosmological constant. \\ 

\section{Results}\label{sec_Results}
\subsection{Analytical Results}
To solve the natural inflationary scenario exactly, we first solve Equation \ref{HJ} for $H$. A potential of the form $V = \left( 3 + A^2/2 \right)B \left( 1 - \cos(A \phi) \right) - BA^2$ has the solution $H = \sqrt{B} \sqrt{1 - \cos A \phi}$. This yields
\begin{equation}\label{epsilone1}
    \begin{aligned}
	\epsilon &=  \frac{A^2}{2} \frac{1 + \cos A \phi}{1 - \cos A \phi}, \\
	\eta &= -A^2/2, \\
	e_1 &= \frac{-A^{-1} + A/2 + (A^{-1} + A/2) \cos A \phi}{\sin A \phi} = \frac{A}{2}\cot A \phi/2 - A^{-1} \tan A \phi/2. \\ 
	\end{aligned}
\end{equation}
Since the potential is cyclical, we restrict ourselves to $\pi > A \phi > 0$. Inflation ends when $e_1 = 0$, which occurs when $\phi_e = A^{-1} \arccos \frac{1 - A^2/2}{1 + A^2/2}$. Starting from the definition of $e_1$ (Equation \ref{E1E2def}), we obtain 
\begin{equation}\label{HTILDEsolve}
	 \tilde{H} \propto exp\left( \int e_1 d \phi \right) \propto \cos(A \phi/2)^{2/A^2} \sin(A \phi/2).
\end{equation}

%However, it is important that $\tilde{H}$ remain positive, and this is not necessarily the case for all interesting regions of $A-\phi$ parameter space, since $A \phi$ may be larger than $\pi/2$. Thus, we instead define 
%\begin{equation}\label{HTILDEsolve}
%	 \tilde{H} \propto |\cos(A \phi/2)|^{2/A^2} |\sin(A \phi/2)|,
%\end{equation}
%which produces the same results as Equations \ref{epsilone1}. \\

We may analyze the $r-n_s$ plane in our model by solving the first of Equations \ref{epsilone1} for $\cos A \phi/2$ and $\sin A \phi/2$ as functions of $\epsilon$ and $A$, which we may use to subsequently solve for $\ln \tilde{H_e}/\tilde{H}$; we obtain
\begin{equation}
	 \tilde{N}_{apprx} = \ln \left\lbrace \frac{1}{2} \left(1 - n_s -A^2 \right)^{-1/A^2} \left( \frac{1 - n_s + A^2}{1 +  A^2/2} \right)^{\frac{2 + A^2}{2A^2}} \right\rbrace.
\end{equation}
Further, we may solve for $r$ explicitly to obtain
\begin{equation}\label{rapprox}
	 r_{apprx} = 2C^{-1} \left\lbrace \left( C \left( 1- n_s \right) + 1 \right)^2 - \left( 1 + CA^2 \right)^2\right\rbrace,
\end{equation}
where in this and the previous equation we have used (from Equation \ref{HTildeCosmoParameters}) the first-order HSRPs for $n_s$, but the second order for $r$. This approximation differs from the numerical results using all first and second order terms by less than about $1 \%$; see Table \ref{Table1}. The deviation of $N$ from $\tilde{N}$ is, from Equation \ref{efolds},
\begin{equation}\label{HChange}
	 \ln H_e/H = \frac{1}{2} \ln \left( \frac{\epsilon + A^2/2}{1 + A^2/2} \right).
\end{equation}

\begin{figure}[]
\begin{subfigure}{.5\textwidth}
  \centering
  \includegraphics[width=1\linewidth]{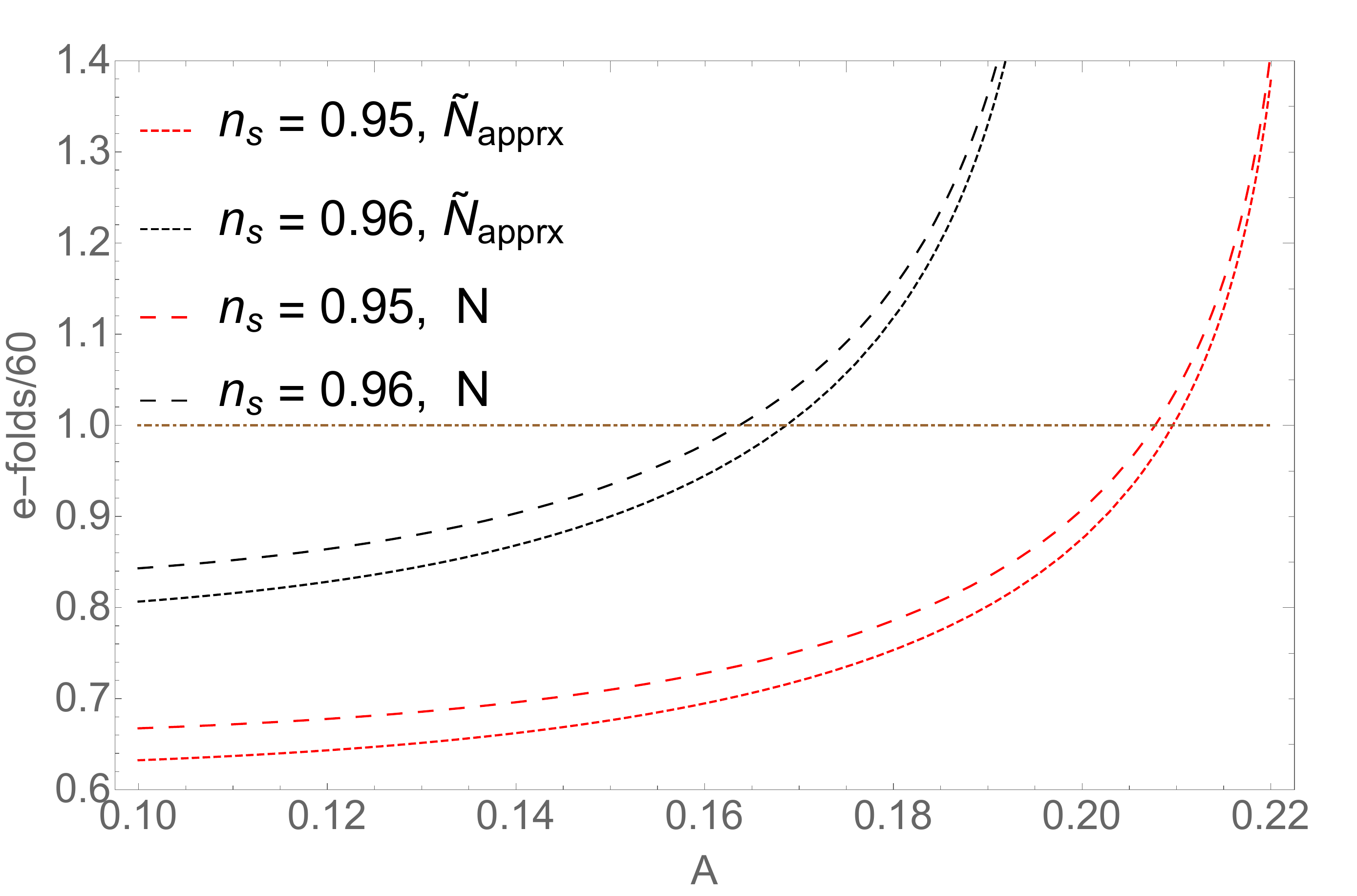}
  \caption{1a}
  \label{fig:analytic1}
\end{subfigure}%
\begin{subfigure}{.5\textwidth}
  \centering
  \includegraphics[width=1\linewidth]{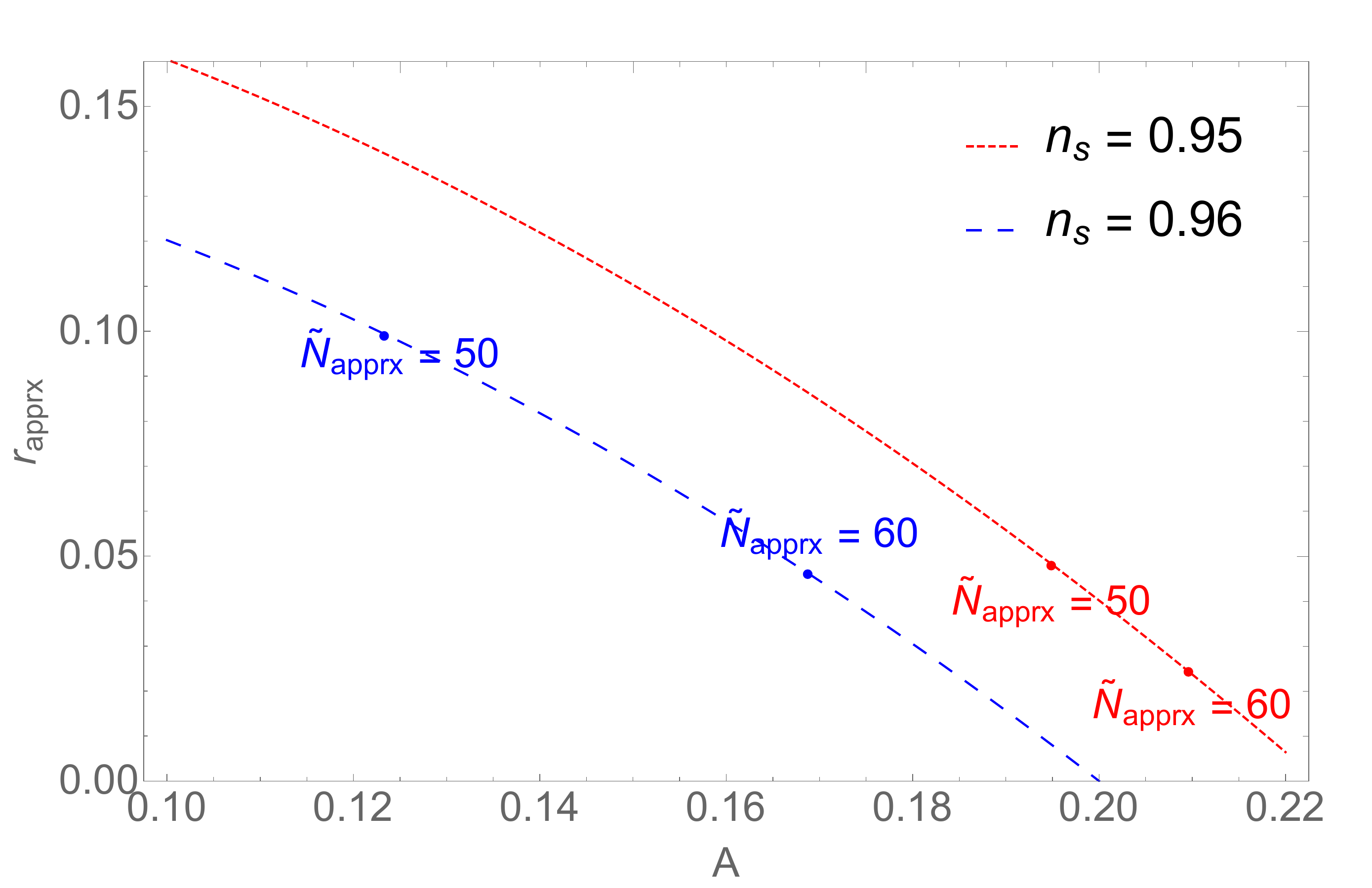}
  \caption{1b}
  \label{fig:analytic2}
\end{subfigure}
\caption{\small In Figure \ref{fig:analytic1}, we show the number of $e$-folds vs $A$ for $\tilde{N}_{apprx}$ and $N$, for two values of $n_s$. In Figure \ref{fig:analytic2}, we plot $r_{apprx}$ vs $A$, specifying 50 and 60 $e$-folds for $n_s = 0.95$ and $0.96$.}
\label{fig:analytic}
\end{figure}

In the slow-roll approximation, the number of $e$-folds is $N \approx \tilde{N}_{apprx} + |\ln H_e/H|$. The latter term has the effect of reducing $A$ when $n_s$ and the number of $e$-folds are kept constant. This is depicted in Figure \ref{fig:analytic1}, where $\tilde{N}_{apprx}$ and $N$ vs. $A$ are plotted for $n_s = 0.96$ and $0.95$. Moreover, a reduction in $A$ reduces the magnitude of the rightmost term of Equation \ref{rapprox}; since this term is negative, however, slow-roll artificially inflates $r$ for a constant $n_s$. This can be seen in Figure \ref{fig:analytic2}, in which we depict $r_{apprx}$ vs $A$ for $n_s = 0.95$ and $0.96$. For $N = 60$, for instance, $\tilde{N}_{apprx} < 60$ due to $\ln H_e/H$. Since smaller $e$-folds correspond to points farther up the curve in Figure \ref{fig:analytic2}, slow-roll overestimates $r$.

%From Equation \ref{HTILDEsolve}, we obtain relations for the cosmological parameters (ref); we have
%\begin{equation}\label{HTildeCosmoParameters}
%    \begin{aligned}
%	\epsilon &= , \\
%	\eta &= -A^2/2, \\
%	\end{aligned}
%\end{equation}

\subsection{Numerical Results}\label{NumResults}
Our numerical results for $n_s$ and $r$ are depicted in Figures \ref{fig:res1} and \ref{fig:res2}. In the latter, our solutions have been computed iteratively, and in the former we have employed parameter scans. In the parameter scans, we have randomly generated values of $\phi$ between $3$ and $33$ and values of $A$ between $0$ and $\pi / \phi$. Only points corresponding to $0.95 < n_s < 0.98$ and $50 >$ $e$-folds $< 70$ are plotted. We apply the bound $2.115 \times 10^{-9} < P_R^{1/2} < 2.315 \times 10^{-9}$, using Equations \ref{CurvPert} and \ref{SRCurvPert} for the exact and slow-roll parameter scan results. 

\begin{figure}[ht]
\begin{subfigure}{.5\textwidth}
  \centering
  \includegraphics[width=1\linewidth]{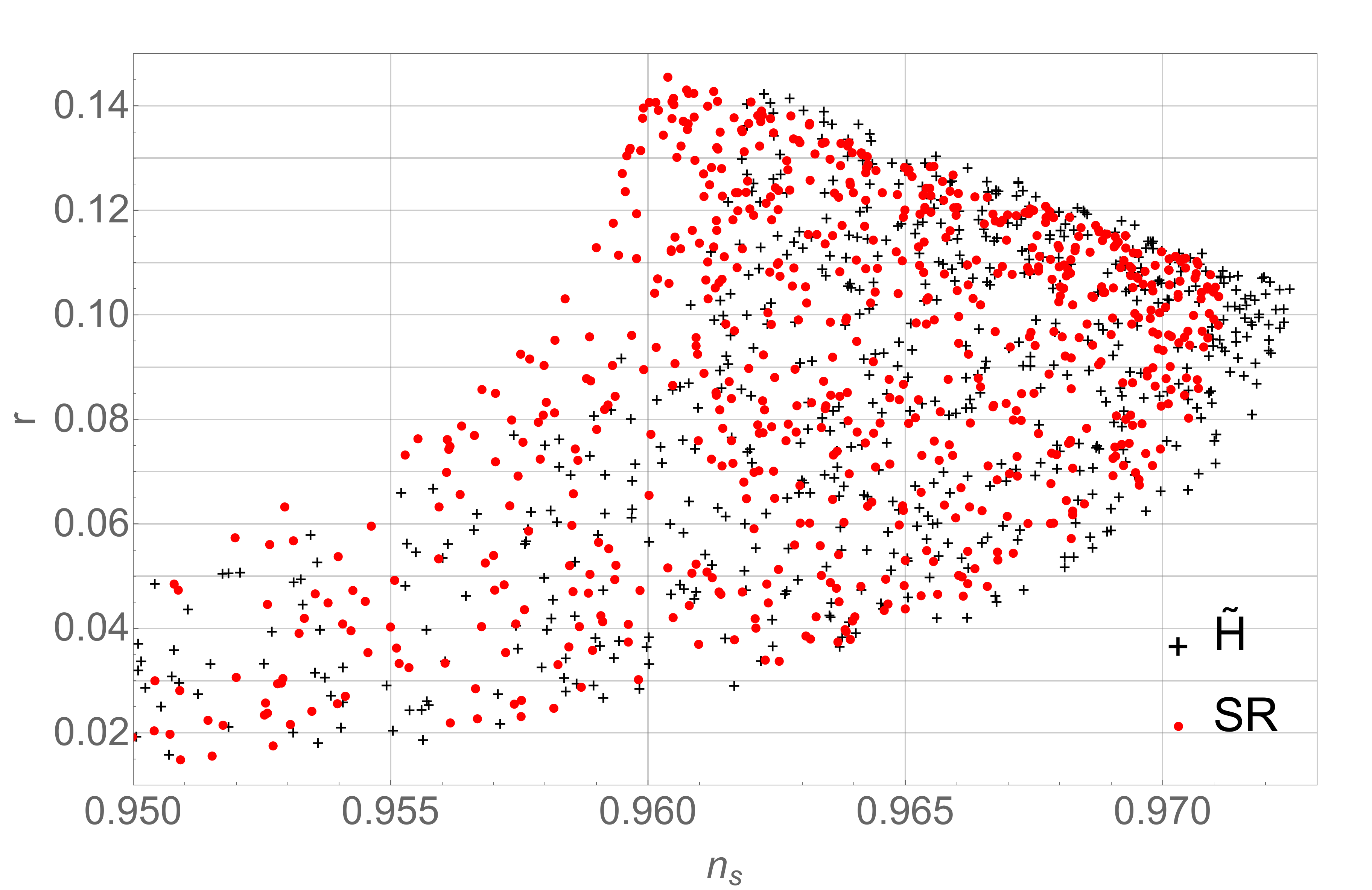}
  \caption{2a}
  \label{fig:res1}
\end{subfigure}%
\begin{subfigure}{.5\textwidth}
  \centering
  \includegraphics[width=1\linewidth]{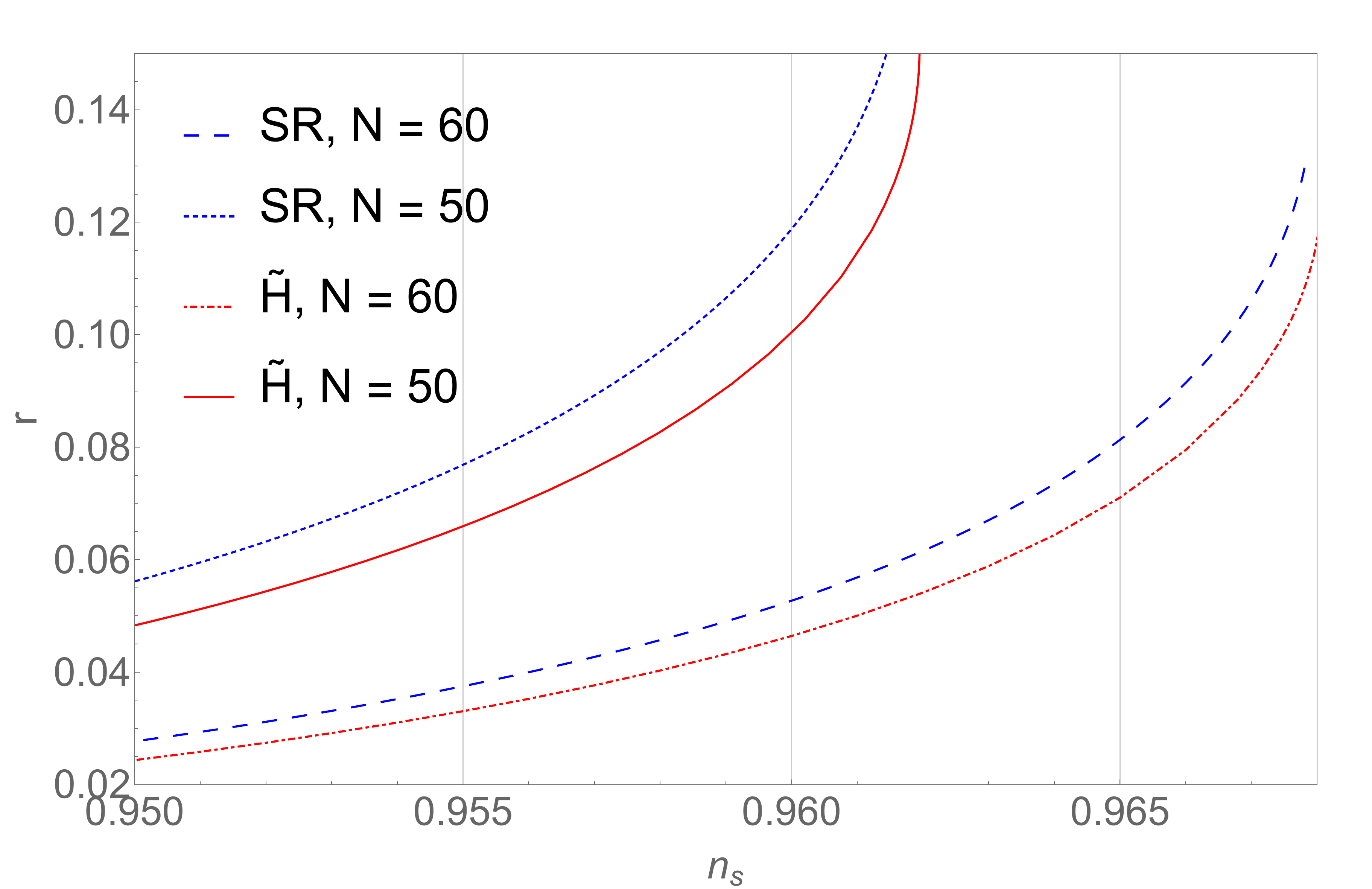}
  \caption{2b}
  \label{fig:res2}
\end{subfigure}
\caption{\small In Figure \ref{fig:res1}, we depict the numerical results in the $r-n_s$ plane for our parameter scan, keeping solutions yielding $50$ to $70$ $e$-folds and curvature perturbations within $2.115 \times 10^{-9} < P_R^{1/2} < 2.315 \times 10^{-9}$. In Figure \ref{fig:res2}, we present numerical results in the $r-n_s$ plane, for $60$ and $50$ $e$-folds during slow-roll (blue dashed and dotted lines, respectively), and $60$ and $50$ $e$-folds without slow-roll (red dot-dashed and solid lines, respectively).}
\label{fig:res}
\end{figure}

\begin{figure}[ht]
\begin{subfigure}{.5\textwidth}
  \centering
  \includegraphics[width=1\linewidth]{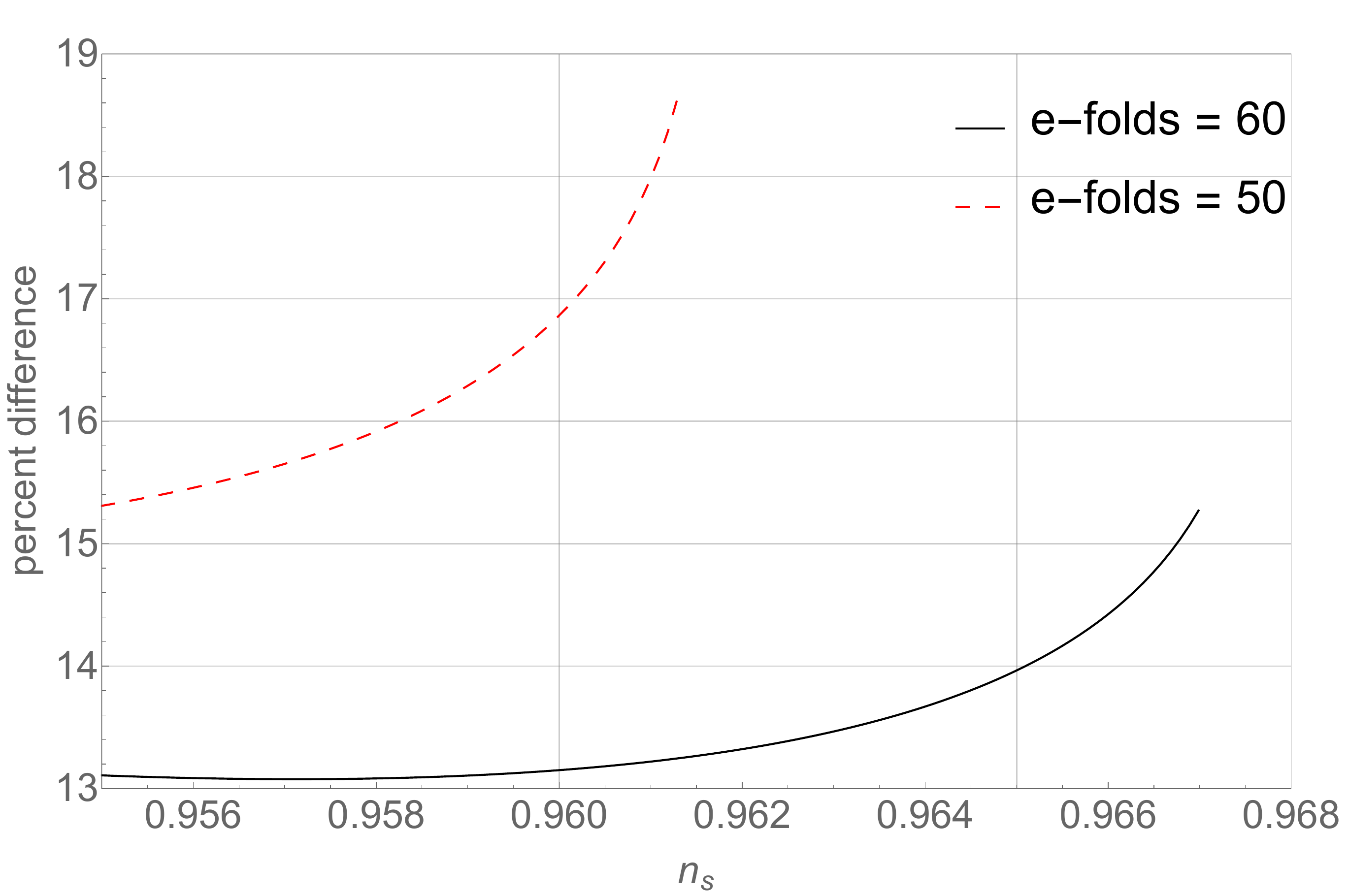}
  \caption{3a}
  \label{fig:err1}
\end{subfigure}%
\begin{subfigure}{.5\textwidth}
  \centering
  \includegraphics[width=1\linewidth]{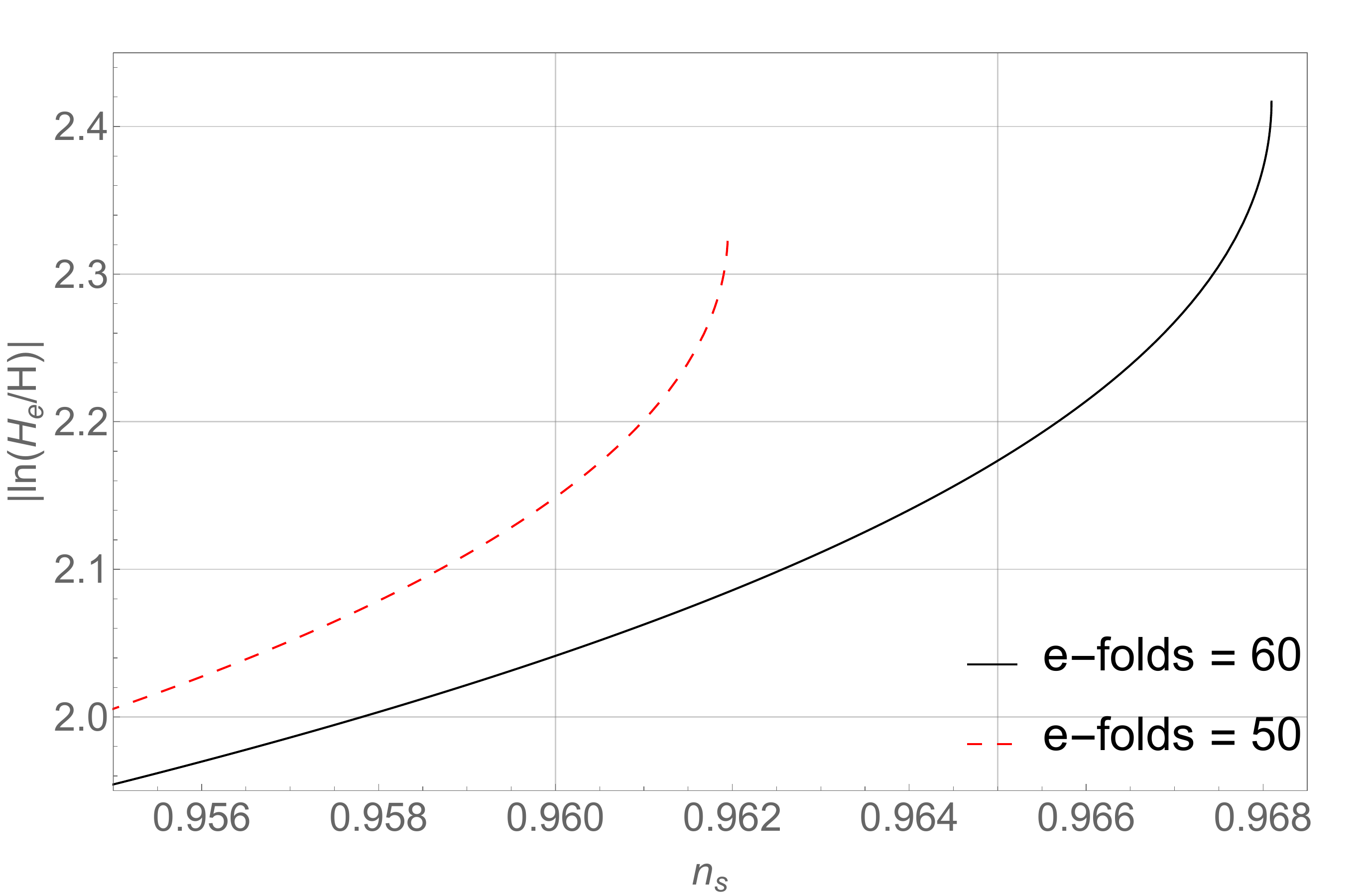}
  \caption{3b}
  \label{fig:err2}
\end{subfigure}
\caption{\small In Figure \ref{fig:err1}, we show the percent difference between the exact and slow-roll solutions. In Figure \ref{fig:err2}, we display $ \ln H_e/H$.}
\label{fig:err}
\end{figure}

\begin{table}[ht]
\begin{tabular}{ |p{1.7cm}||p{1.7cm}|p{1.7cm}|p{1.7cm}|p{1.7cm}|p{1.1cm}|p{1.1cm}|p{1.1cm}|p{1.1cm}|p{0.8cm}|  }
 \hline
% \multicolumn{5}{|c|}{} \\
 \hline
 $n_s$ &$r$ & $r_{SR}$ & $r_{apprx}$ & $ \ln H_e/H$ & $\%$ diff  \\
 \hline
% $0.949987$   & 0.0241324  &  0.0275903 & 0.0241444 &  -1.8865 & 13.3709  \\\hline
% $0.952487$ &   0.0280481 & 0.0320159 & 0.02806  & -1.9185 & 13.2118  \\\hline
% $0.954987$ & 0.0328488 & 0.035789 & 0.0328021 &  -1.9545 & 13.1086  \\\hline
 $0.957487$ & 0.038652 & 0.0440606 & 0.0388604 &  -1.99435 & 13.0781  \\\hline
 $0.959987$ & 0.0461037  & 0.0525931 & 0.046308 & -2.04113 & 13.1501  \\\hline
 $0.962487$ & 0.0560409  & 0.0640809 & 0.0562123   & -2.09795 & 13.3864  \\\hline
 $0.964987$ & 0.0705856  & 0.0811792 & 0.0706222 & -2.17312 & 13.9605  \\\hline
 $0.967487$ & 0.09936  & 0.116638 & 0.0982739 & -2.30444 & 15.9982  \\
 \hline
\end{tabular}
 \caption{\small Here we tabulate $n_s$, $r$, $r_{SR}$, $r_{apprx}$, $\ln H_e/H$, and the percent difference between the exact results ($r$) and the slow-roll results ($r_{SR}$). We assume $60$ $e$-folds throughout.}\label{Table1}
\label{table}
\end{table}

\indent We have mentioned that slow-roll overestimates $r$ for a constant $n_s$ and $e$-folds. Figure \ref{fig:res1}, however, indicates that slow-roll also artificially drags the solutions leftward along the $n_s$ axis. A larger proportion of the exact solutions compared to slow roll corresponds to $n_s > 0.965$; in particular, in Figure \ref{fig:res1} about $52\%$ and $38\%$ of the solutions are $n_s > 0.965$, for $\tilde{H}$ and SR respectively. Further, about $83\%$ and $72\%$ of the solutions are $n_s > 0.960$, for $\tilde{H}$ and SR respectively. This is a significant improvement in light of the Planck 2018 results (see \cite{aghanim2018planck}), which indicate that $n_s =  0.9665 \pm 0.0038$ for TT, TE, EE + low E + lensing + BAO. \\
\indent In Figures \ref{fig:err1} and \ref{fig:err2}, our iterative results for percent difference and $\ln H_e/H$ are shown. The former directly quantifies the error introduced by slow-roll, and the latter quantifies the deviation of $\epsilon$ from $1$. Here we note that slow-roll artificially inflates $r$ by about $13-19 \%$ for $50-60$ $e$-folds, around $n_s \approx 0.96$. \\
\indent The running of the spectral index is negligible, and slow-roll does not significantly affect it in light of its experimental bounds. Finally, in this paper we have assumed instantaneous reheating. Reheating in natural inflation, however, was discussed in a recent paper \cite{wu2018constraints}. The cited paper demonstrates that reheating is generally insensitive to the value of $A$, even in the case of two-phase reheating. It may be interesting, however, to investigate reheating in the context of natural $e$-folds in subsequent work.

\section{Conclusion}\label{conclusion}
We have examined a natural inflationary scenario without the use of the slow-roll approximation, via the Hamilton-Jacobi equation and a reparametrization from $V$ to $\tilde{H} = aH$. This method allows one to solve for $r$ and $n_s$ without slow-roll, and to compute the number of $e$-folds by simply evaluating $\ln (\tilde{H}_e / \tilde{H} )$. Thus, we avoid the typically onerous numerical integration techniques required by slow-roll. In doing so, we show that at around $n_s \approx 0.96$, slow-roll overestimates $r$ by $13-19 \%$ for $50-60$ $e$-folds. As cosmological data are expected to become more precise in the future, we expect that this error may become significant. \\
\indent We have also noted that this reparametrization is difficult to implement in general, given the complexity of Equations \ref{HJ} and \ref{HTILDEfromH}. It would therefore be useful to consider a numerical generalization of the exact techniques presented here, which may allow for the numerical simulation of the HSRPs from $V$, and hence the mitigation of the errors produced by slow-roll.

\begin{acknowledgments}
Support for this project was provided by a PSC-CUNY Award, jointly funded by The Professional Staff Congress and The City University of New York.
\end{acknowledgments}

%%%%%		Appendix I
\appendix

\renewcommand{\thesection}{\arabic{section}}
\renewcommand{\thesubsection}{\arabic{subsection}}
%\section{Acknowledgments}
%
%Support for this project was provided by a PSC-CUNY Award, jointly funded by The Professional Staff Congress and The City University of New York.

%\section{$e$ parameters}\label{app1}
%The parameters $e_2$ and $e_3$ are, from Equation \ref{E1E2def},
%\begin{equation}\label{epsilone1Natural}
%    \begin{aligned}
%	e_2 &= -\frac{A^2 (6 + A^2) + 2 (-2 + A^2) \tan^2 A \phi/2}{4 A^2}, \\
%	e_3 &= -\frac{A^4 (6 + A^2) \cot^2 A \phi/2 + 4 A^2 (-6 + A^2) \tan A \phi/2 + 4(2 - 3 A^2 + A^4) \tan^3 A \phi/2}{8 A^3}  
%	\end{aligned}
%\end{equation}
%%%%%		Appendix II

\bibliography{NaturalInflation}

%merlin.mbs apsrev4-1.bst 2010-07-25 4.21a (PWD, AO, DPC) hacked
%Control: key (0)
%Control: author (8) initials jnrlst
%Control: editor formatted (1) identically to author
%Control: production of article title (-1) disabled
%Control: page (0) single
%Control: year (1) truncated
%Control: production of eprint (0) enabled
\begin{thebibliography}{22}%
\makeatletter
\providecommand \@ifxundefined [1]{%
 \@ifx{#1\undefined}
}%
\providecommand \@ifnum [1]{%
 \ifnum #1\expandafter \@firstoftwo
 \else \expandafter \@secondoftwo
 \fi
}%
\providecommand \@ifx [1]{%
 \ifx #1\expandafter \@firstoftwo
 \else \expandafter \@secondoftwo
 \fi
}%
\providecommand \natexlab [1]{#1}%
\providecommand \enquote  [1]{``#1''}%
\providecommand \bibnamefont  [1]{#1}%
\providecommand \bibfnamefont [1]{#1}%
\providecommand \citenamefont [1]{#1}%
\providecommand \href@noop [0]{\@secondoftwo}%
\providecommand \href [0]{\begingroup \@sanitize@url \@href}%
\providecommand \@href[1]{\@@startlink{#1}\@@href}%
\providecommand \@@href[1]{\endgroup#1\@@endlink}%
\providecommand \@sanitize@url [0]{\catcode `\\12\catcode `\$12\catcode
  `\&12\catcode `\#12\catcode `\^12\catcode `\_12\catcode `\%12\relax}%
\providecommand \@@startlink[1]{}%
\providecommand \@@endlink[0]{}%
\providecommand \url  [0]{\begingroup\@sanitize@url \@url }%
\providecommand \@url [1]{\endgroup\@href {#1}{\urlprefix }}%
\providecommand \urlprefix  [0]{URL }%
\providecommand \Eprint [0]{\href }%
\providecommand \doibase [0]{http://dx.doi.org/}%
\providecommand \selectlanguage [0]{\@gobble}%
\providecommand \bibinfo  [0]{\@secondoftwo}%
\providecommand \bibfield  [0]{\@secondoftwo}%
\providecommand \translation [1]{[#1]}%
\providecommand \BibitemOpen [0]{}%
\providecommand \bibitemStop [0]{}%
\providecommand \bibitemNoStop [0]{.\EOS\space}%
\providecommand \EOS [0]{\spacefactor3000\relax}%
\providecommand \BibitemShut  [1]{\csname bibitem#1\endcsname}%
\let\auto@bib@innerbib\@empty
%</preamble>
\bibitem [{\citenamefont {Guth}(1981)}]{guth1981inflationary}%
  \BibitemOpen
  \bibfield  {author} {\bibinfo {author} {\bibfnamefont {A.~H.}\ \bibnamefont
  {Guth}},\ }\href@noop {} {\bibfield  {journal} {\bibinfo  {journal} {Physical
  Review D}\ }\textbf {\bibinfo {volume} {23}},\ \bibinfo {pages} {347}
  (\bibinfo {year} {1981})}\BibitemShut {NoStop}%
\bibitem [{\citenamefont {Bennett}\ \emph {et~al.}(2013)\citenamefont
  {Bennett}, \citenamefont {Larson}, \citenamefont {Weiland}, \citenamefont
  {Jarosik}, \citenamefont {Hinshaw}, \citenamefont {Odegard}, \citenamefont
  {Smith}, \citenamefont {Hill}, \citenamefont {Gold}, \citenamefont {Halpern}
  \emph {et~al.}}]{bennett2013nine}%
  \BibitemOpen
  \bibfield  {author} {\bibinfo {author} {\bibfnamefont {C.}~\bibnamefont
  {Bennett}}, \bibinfo {author} {\bibfnamefont {D.}~\bibnamefont {Larson}},
  \bibinfo {author} {\bibfnamefont {J.}~\bibnamefont {Weiland}}, \bibinfo
  {author} {\bibfnamefont {N.}~\bibnamefont {Jarosik}}, \bibinfo {author}
  {\bibfnamefont {G.}~\bibnamefont {Hinshaw}}, \bibinfo {author} {\bibfnamefont
  {N.}~\bibnamefont {Odegard}}, \bibinfo {author} {\bibfnamefont
  {K.}~\bibnamefont {Smith}}, \bibinfo {author} {\bibfnamefont
  {R.}~\bibnamefont {Hill}}, \bibinfo {author} {\bibfnamefont {B.}~\bibnamefont
  {Gold}}, \bibinfo {author} {\bibfnamefont {M.}~\bibnamefont {Halpern}},
  \emph {et~al.},\ }\href@noop {} {\bibfield  {journal} {\bibinfo  {journal}
  {The Astrophysical Journal Supplement Series}\ }\textbf {\bibinfo {volume}
  {208}},\ \bibinfo {pages} {20} (\bibinfo {year} {2013})}\BibitemShut
  {NoStop}%
\bibitem [{\citenamefont {Ade}\ \emph {et~al.}(2016)\citenamefont {Ade},
  \citenamefont {Aghanim}, \citenamefont {Arnaud}, \citenamefont {Ashdown},
  \citenamefont {Aumont}, \citenamefont {Baccigalupi}, \citenamefont {Banday},
  \citenamefont {Barreiro}, \citenamefont {Bartlett}, \citenamefont {Bartolo}
  \emph {et~al.}}]{ade2016planck}%
  \BibitemOpen
  \bibfield  {author} {\bibinfo {author} {\bibfnamefont {P.~A.}\ \bibnamefont
  {Ade}}, \bibinfo {author} {\bibfnamefont {N.}~\bibnamefont {Aghanim}},
  \bibinfo {author} {\bibfnamefont {M.}~\bibnamefont {Arnaud}}, \bibinfo
  {author} {\bibfnamefont {M.}~\bibnamefont {Ashdown}}, \bibinfo {author}
  {\bibfnamefont {J.}~\bibnamefont {Aumont}}, \bibinfo {author} {\bibfnamefont
  {C.}~\bibnamefont {Baccigalupi}}, \bibinfo {author} {\bibfnamefont
  {A.}~\bibnamefont {Banday}}, \bibinfo {author} {\bibfnamefont
  {R.}~\bibnamefont {Barreiro}}, \bibinfo {author} {\bibfnamefont
  {J.}~\bibnamefont {Bartlett}}, \bibinfo {author} {\bibfnamefont
  {N.}~\bibnamefont {Bartolo}},  \emph {et~al.},\ }\href@noop {} {\bibfield
  {journal} {\bibinfo  {journal} {Astronomy \& Astrophysics}\ }\textbf
  {\bibinfo {volume} {594}},\ \bibinfo {pages} {A13} (\bibinfo {year}
  {2016})}\BibitemShut {NoStop}%
\bibitem [{\citenamefont {Hazumi}\ \emph {et~al.}(2019)\citenamefont {Hazumi},
  \citenamefont {Ade}, \citenamefont {Akiba}, \citenamefont {Alonso},
  \citenamefont {Arnold}, \citenamefont {Aumont}, \citenamefont {Baccigalupi},
  \citenamefont {Barron}, \citenamefont {Basak}, \citenamefont {Beckman} \emph
  {et~al.}}]{hazumi2019litebird}%
  \BibitemOpen
  \bibfield  {author} {\bibinfo {author} {\bibfnamefont {M.}~\bibnamefont
  {Hazumi}}, \bibinfo {author} {\bibfnamefont {P.}~\bibnamefont {Ade}},
  \bibinfo {author} {\bibfnamefont {Y.}~\bibnamefont {Akiba}}, \bibinfo
  {author} {\bibfnamefont {D.}~\bibnamefont {Alonso}}, \bibinfo {author}
  {\bibfnamefont {K.}~\bibnamefont {Arnold}}, \bibinfo {author} {\bibfnamefont
  {J.}~\bibnamefont {Aumont}}, \bibinfo {author} {\bibfnamefont
  {C.}~\bibnamefont {Baccigalupi}}, \bibinfo {author} {\bibfnamefont
  {D.}~\bibnamefont {Barron}}, \bibinfo {author} {\bibfnamefont
  {S.}~\bibnamefont {Basak}}, \bibinfo {author} {\bibfnamefont
  {S.}~\bibnamefont {Beckman}},  \emph {et~al.},\ }\href@noop {} {\bibfield
  {journal} {\bibinfo  {journal} {Journal of Low Temperature Physics}\ }\textbf
  {\bibinfo {volume} {194}},\ \bibinfo {pages} {443} (\bibinfo {year}
  {2019})}\BibitemShut {NoStop}%
\bibitem [{\citenamefont {Tartari}\ \emph {et~al.}(2016)\citenamefont
  {Tartari}, \citenamefont {Aumont}, \citenamefont {Banfi}, \citenamefont
  {Battaglia}, \citenamefont {Battistelli}, \citenamefont {Ba{\`u}},
  \citenamefont {B{\'e}lier}, \citenamefont {Bennett}, \citenamefont
  {Berg{\'e}}, \citenamefont {Bernard} \emph {et~al.}}]{tartari2016qubic}%
  \BibitemOpen
  \bibfield  {author} {\bibinfo {author} {\bibfnamefont {A.}~\bibnamefont
  {Tartari}}, \bibinfo {author} {\bibfnamefont {J.}~\bibnamefont {Aumont}},
  \bibinfo {author} {\bibfnamefont {S.}~\bibnamefont {Banfi}}, \bibinfo
  {author} {\bibfnamefont {P.}~\bibnamefont {Battaglia}}, \bibinfo {author}
  {\bibfnamefont {E.}~\bibnamefont {Battistelli}}, \bibinfo {author}
  {\bibfnamefont {A.}~\bibnamefont {Ba{\`u}}}, \bibinfo {author} {\bibfnamefont
  {B.}~\bibnamefont {B{\'e}lier}}, \bibinfo {author} {\bibfnamefont
  {D.}~\bibnamefont {Bennett}}, \bibinfo {author} {\bibfnamefont
  {L.}~\bibnamefont {Berg{\'e}}}, \bibinfo {author} {\bibfnamefont {J.~P.}\
  \bibnamefont {Bernard}},  \emph {et~al.},\ }\href@noop {} {\bibfield
  {journal} {\bibinfo  {journal} {Journal of Low Temperature Physics}\ }\textbf
  {\bibinfo {volume} {184}},\ \bibinfo {pages} {739} (\bibinfo {year}
  {2016})}\BibitemShut {NoStop}%
\bibitem [{\citenamefont {Freese}\ \emph {et~al.}(1990)\citenamefont {Freese},
  \citenamefont {Frieman},\ and\ \citenamefont {Olinto}}]{freese1990natural}%
  \BibitemOpen
  \bibfield  {author} {\bibinfo {author} {\bibfnamefont {K.}~\bibnamefont
  {Freese}}, \bibinfo {author} {\bibfnamefont {J.~A.}\ \bibnamefont {Frieman}},
  \ and\ \bibinfo {author} {\bibfnamefont {A.~V.}\ \bibnamefont {Olinto}},\
  }\href@noop {} {\bibfield  {journal} {\bibinfo  {journal} {Physical Review
  Letters}\ }\textbf {\bibinfo {volume} {65}},\ \bibinfo {pages} {3233}
  (\bibinfo {year} {1990})}\BibitemShut {NoStop}%
\bibitem [{\citenamefont {Peccei}\ and\ \citenamefont
  {Quinn}(1977)}]{peccei1977cp}%
  \BibitemOpen
  \bibfield  {author} {\bibinfo {author} {\bibfnamefont {R.~D.}\ \bibnamefont
  {Peccei}}\ and\ \bibinfo {author} {\bibfnamefont {H.~R.}\ \bibnamefont
  {Quinn}},\ }\href@noop {} {\bibfield  {journal} {\bibinfo  {journal}
  {Physical Review Letters}\ }\textbf {\bibinfo {volume} {38}},\ \bibinfo
  {pages} {1440} (\bibinfo {year} {1977})}\BibitemShut {NoStop}%
\bibitem [{\citenamefont {Freese}\ and\ \citenamefont
  {Kinney}(2015)}]{freese2015natural}%
  \BibitemOpen
  \bibfield  {author} {\bibinfo {author} {\bibfnamefont {K.}~\bibnamefont
  {Freese}}\ and\ \bibinfo {author} {\bibfnamefont {W.~H.}\ \bibnamefont
  {Kinney}},\ }\href@noop {} {\bibfield  {journal} {\bibinfo  {journal}
  {Journal of Cosmology and Astroparticle Physics}\ }\textbf {\bibinfo {volume}
  {2015}},\ \bibinfo {pages} {044} (\bibinfo {year} {2015})}\BibitemShut
  {NoStop}%
\bibitem [{\citenamefont {Ross}\ \emph {et~al.}(2016)\citenamefont {Ross},
  \citenamefont {Germ{\'a}n},\ and\ \citenamefont
  {V{\'a}zquez}}]{ross2016hybrid}%
  \BibitemOpen
  \bibfield  {author} {\bibinfo {author} {\bibfnamefont {G.~G.}\ \bibnamefont
  {Ross}}, \bibinfo {author} {\bibfnamefont {G.}~\bibnamefont {Germ{\'a}n}}, \
  and\ \bibinfo {author} {\bibfnamefont {J.~A.}\ \bibnamefont {V{\'a}zquez}},\
  }\href@noop {} {\bibfield  {journal} {\bibinfo  {journal} {Journal of High
  Energy Physics}\ }\textbf {\bibinfo {volume} {2016}},\ \bibinfo {pages} {10}
  (\bibinfo {year} {2016})}\BibitemShut {NoStop}%
\bibitem [{\citenamefont {Ross}\ and\ \citenamefont
  {Germ{\'a}n}(2010{\natexlab{a}})}]{ross2010hybrid}%
  \BibitemOpen
  \bibfield  {author} {\bibinfo {author} {\bibfnamefont {G.~G.}\ \bibnamefont
  {Ross}}\ and\ \bibinfo {author} {\bibfnamefont {G.}~\bibnamefont
  {Germ{\'a}n}},\ }\href@noop {} {\bibfield  {journal} {\bibinfo  {journal}
  {Physics Letters B}\ }\textbf {\bibinfo {volume} {684}},\ \bibinfo {pages}
  {199} (\bibinfo {year} {2010}{\natexlab{a}})}\BibitemShut {NoStop}%
\bibitem [{\citenamefont {Ross}\ and\ \citenamefont
  {Germ{\'a}n}(2010{\natexlab{b}})}]{ross2010hybrid2}%
  \BibitemOpen
  \bibfield  {author} {\bibinfo {author} {\bibfnamefont {G.~G.}\ \bibnamefont
  {Ross}}\ and\ \bibinfo {author} {\bibfnamefont {G.}~\bibnamefont
  {Germ{\'a}n}},\ }\href@noop {} {\bibfield  {journal} {\bibinfo  {journal}
  {Physics Letters B}\ }\textbf {\bibinfo {volume} {691}},\ \bibinfo {pages}
  {117} (\bibinfo {year} {2010}{\natexlab{b}})}\BibitemShut {NoStop}%
\bibitem [{\citenamefont {Germ{\'a}n}\ \emph {et~al.}(2017)\citenamefont
  {Germ{\'a}n}, \citenamefont {Herrera-Aguilar}, \citenamefont {Hidalgo},
  \citenamefont {Sussman},\ and\ \citenamefont {Tapia}}]{german2017general}%
  \BibitemOpen
  \bibfield  {author} {\bibinfo {author} {\bibfnamefont {G.}~\bibnamefont
  {Germ{\'a}n}}, \bibinfo {author} {\bibfnamefont {A.}~\bibnamefont
  {Herrera-Aguilar}}, \bibinfo {author} {\bibfnamefont {J.~C.}\ \bibnamefont
  {Hidalgo}}, \bibinfo {author} {\bibfnamefont {R.~A.}\ \bibnamefont
  {Sussman}}, \ and\ \bibinfo {author} {\bibfnamefont {J.}~\bibnamefont
  {Tapia}},\ }\href@noop {} {\bibfield  {journal} {\bibinfo  {journal} {Journal
  of Cosmology and Astroparticle Physics}\ }\textbf {\bibinfo {volume}
  {2017}},\ \bibinfo {pages} {003} (\bibinfo {year} {2017})}\BibitemShut
  {NoStop}%
\bibitem [{\citenamefont {Motohashi}\ \emph {et~al.}(2015)\citenamefont
  {Motohashi}, \citenamefont {Starobinsky},\ and\ \citenamefont
  {Yokoyama}}]{motohashi2015inflation}%
  \BibitemOpen
  \bibfield  {author} {\bibinfo {author} {\bibfnamefont {H.}~\bibnamefont
  {Motohashi}}, \bibinfo {author} {\bibfnamefont {A.~A.}\ \bibnamefont
  {Starobinsky}}, \ and\ \bibinfo {author} {\bibfnamefont {J.}~\bibnamefont
  {Yokoyama}},\ }\href@noop {} {\bibfield  {journal} {\bibinfo  {journal}
  {Journal of Cosmology and Astroparticle Physics}\ }\textbf {\bibinfo {volume}
  {2015}},\ \bibinfo {pages} {018} (\bibinfo {year} {2015})}\BibitemShut
  {NoStop}%
\bibitem [{\citenamefont {Mithani}\ and\ \citenamefont
  {Vilenkin}(2013)}]{mithani2013inflation}%
  \BibitemOpen
  \bibfield  {author} {\bibinfo {author} {\bibfnamefont {A.~T.}\ \bibnamefont
  {Mithani}}\ and\ \bibinfo {author} {\bibfnamefont {A.}~\bibnamefont
  {Vilenkin}},\ }\href@noop {} {\bibfield  {journal} {\bibinfo  {journal}
  {Journal of Cosmology and Astroparticle Physics}\ }\textbf {\bibinfo {volume}
  {2013}},\ \bibinfo {pages} {024} (\bibinfo {year} {2013})}\BibitemShut
  {NoStop}%
\bibitem [{\citenamefont {Liddle}\ \emph {et~al.}(1994)\citenamefont {Liddle},
  \citenamefont {Parsons},\ and\ \citenamefont
  {Barrow}}]{liddle1994formalizing}%
  \BibitemOpen
  \bibfield  {author} {\bibinfo {author} {\bibfnamefont {A.~R.}\ \bibnamefont
  {Liddle}}, \bibinfo {author} {\bibfnamefont {P.}~\bibnamefont {Parsons}}, \
  and\ \bibinfo {author} {\bibfnamefont {J.~D.}\ \bibnamefont {Barrow}},\
  }\href@noop {} {\bibfield  {journal} {\bibinfo  {journal} {Physical Review
  D}\ }\textbf {\bibinfo {volume} {50}},\ \bibinfo {pages} {7222} (\bibinfo
  {year} {1994})}\BibitemShut {NoStop}%
\bibitem [{\citenamefont {Stewart}\ and\ \citenamefont
  {Lyth}(1993)}]{stewart1993more}%
  \BibitemOpen
  \bibfield  {author} {\bibinfo {author} {\bibfnamefont {E.~D.}\ \bibnamefont
  {Stewart}}\ and\ \bibinfo {author} {\bibfnamefont {D.~H.}\ \bibnamefont
  {Lyth}},\ }\href@noop {} {\bibfield  {journal} {\bibinfo  {journal} {Phys.
  Lett.}\ }\textbf {\bibinfo {volume} {302}},\ \bibinfo {pages} {171} (\bibinfo
  {year} {1993})}\BibitemShut {NoStop}%
\bibitem [{\citenamefont {Lidsey}\ \emph {et~al.}(1997)\citenamefont {Lidsey},
  \citenamefont {Liddle}, \citenamefont {Kolb}, \citenamefont {Copeland},
  \citenamefont {Barreiro},\ and\ \citenamefont
  {Abney}}]{lidsey1997reconstructing}%
  \BibitemOpen
  \bibfield  {author} {\bibinfo {author} {\bibfnamefont {J.~E.}\ \bibnamefont
  {Lidsey}}, \bibinfo {author} {\bibfnamefont {A.~R.}\ \bibnamefont {Liddle}},
  \bibinfo {author} {\bibfnamefont {E.~W.}\ \bibnamefont {Kolb}}, \bibinfo
  {author} {\bibfnamefont {E.~J.}\ \bibnamefont {Copeland}}, \bibinfo {author}
  {\bibfnamefont {T.}~\bibnamefont {Barreiro}}, \ and\ \bibinfo {author}
  {\bibfnamefont {M.}~\bibnamefont {Abney}},\ }\href@noop {} {\bibfield
  {journal} {\bibinfo  {journal} {Reviews of Modern Physics}\ }\textbf
  {\bibinfo {volume} {69}},\ \bibinfo {pages} {373} (\bibinfo {year}
  {1997})}\BibitemShut {NoStop}%
\bibitem [{\citenamefont {Chongchitnan}(2016)}]{chongchitnan2016inflation}%
  \BibitemOpen
  \bibfield  {author} {\bibinfo {author} {\bibfnamefont {S.}~\bibnamefont
  {Chongchitnan}},\ }\href@noop {} {\bibfield  {journal} {\bibinfo  {journal}
  {Physical Review D}\ }\textbf {\bibinfo {volume} {94}},\ \bibinfo {pages}
  {043526} (\bibinfo {year} {2016})}\BibitemShut {NoStop}%
\bibitem [{\citenamefont
  {Chongchitnan}(2017{\natexlab{a}})}]{chongchitnan2017inflationary}%
  \BibitemOpen
  \bibfield  {author} {\bibinfo {author} {\bibfnamefont {S.}~\bibnamefont
  {Chongchitnan}},\ }\href@noop {} {\bibfield  {journal} {\bibinfo  {journal}
  {arXiv preprint arXiv:1705.02712}\ } (\bibinfo {year}
  {2017}{\natexlab{a}})}\BibitemShut {NoStop}%
\bibitem [{\citenamefont
  {Chongchitnan}(2017{\natexlab{b}})}]{chongchitnan2017reheating}%
  \BibitemOpen
  \bibfield  {author} {\bibinfo {author} {\bibfnamefont {S.}~\bibnamefont
  {Chongchitnan}},\ }\href@noop {} {\bibfield  {journal} {\bibinfo  {journal}
  {arXiv preprint arXiv:1709.03482}\ } (\bibinfo {year}
  {2017}{\natexlab{b}})}\BibitemShut {NoStop}%
\bibitem [{\citenamefont {Aghanim}\ \emph {et~al.}(2018)\citenamefont
  {Aghanim}, \citenamefont {Polastri}, \citenamefont {Rubi{\~n}o-Mart{\'\i}n},
  \citenamefont {Dupac}, \citenamefont {Liguori}, \citenamefont {Kim},
  \citenamefont {Matarrese}, \citenamefont {G{\'e}nova-Santos}, \citenamefont
  {Huang}, \citenamefont {Forastieri} \emph {et~al.}}]{aghanim2018planck}%
  \BibitemOpen
  \bibfield  {author} {\bibinfo {author} {\bibfnamefont {N.}~\bibnamefont
  {Aghanim}}, \bibinfo {author} {\bibfnamefont {L.}~\bibnamefont {Polastri}},
  \bibinfo {author} {\bibfnamefont {J.}~\bibnamefont {Rubi{\~n}o-Mart{\'\i}n}},
  \bibinfo {author} {\bibfnamefont {X.}~\bibnamefont {Dupac}}, \bibinfo
  {author} {\bibfnamefont {M.}~\bibnamefont {Liguori}}, \bibinfo {author}
  {\bibfnamefont {J.}~\bibnamefont {Kim}}, \bibinfo {author} {\bibfnamefont
  {S.}~\bibnamefont {Matarrese}}, \bibinfo {author} {\bibfnamefont
  {R.}~\bibnamefont {G{\'e}nova-Santos}}, \bibinfo {author} {\bibfnamefont
  {Z.}~\bibnamefont {Huang}}, \bibinfo {author} {\bibfnamefont
  {F.}~\bibnamefont {Forastieri}},  \emph {et~al.},\ }\href@noop {} {\emph
  {\bibinfo {title} {Planck 2018 results. VI. Cosmological parameters}}},\
  \bibinfo {type} {Tech. Rep.}\ (\bibinfo {year} {2018})\BibitemShut {NoStop}%
\bibitem [{\citenamefont {Wu}\ \emph {et~al.}(2018)\citenamefont {Wu},
  \citenamefont {Zhang}, \citenamefont {Sun}, \citenamefont {Shou},\ and\
  \citenamefont {Xu}}]{wu2018constraints}%
  \BibitemOpen
  \bibfield  {author} {\bibinfo {author} {\bibfnamefont {Y.-B.}\ \bibnamefont
  {Wu}}, \bibinfo {author} {\bibfnamefont {N.}~\bibnamefont {Zhang}}, \bibinfo
  {author} {\bibfnamefont {C.-W.}\ \bibnamefont {Sun}}, \bibinfo {author}
  {\bibfnamefont {L.-J.}\ \bibnamefont {Shou}}, \ and\ \bibinfo {author}
  {\bibfnamefont {H.-Z.}\ \bibnamefont {Xu}},\ }\href@noop {} {\bibfield
  {journal} {\bibinfo  {journal} {arXiv preprint arXiv:1807.03596}\ } (\bibinfo
  {year} {2018})}\BibitemShut {NoStop}%
\end{thebibliography}%

\end{document}